\documentclass[twocolumn,showpacs,fleqn,nobibnotes]{revtex4}

\usepackage{amsmath}
\usepackage{graphicx}
\usepackage{float}
\usepackage{subfigure}

\def\lsim{\raise0.3ex\hbox{$<$\kern-0.75em\raise-1.1ex\hbox{$\sim$}}}
\def\gsim{\raise0.3ex\hbox{$>$\kern-0.75em\raise-1.1ex\hbox{$\sim$}}}

\renewcommand{\vec}[1]{\boldsymbol{#1}}
\newcommand{\dif}{\mathrm{d}}

\begin{document}

\title{High energy electroweak DVCS processes in neutrino-nucleon scattering}
\author{Magno V.T. Machado}
\affiliation{Centro de Ci\^encias Exatas e Tecnol\'ogicas, Universidade Federal do Pampa \\
Campus de Bag\'e, Rua Carlos Barbosa. CEP 96400-970. Bag\'e, RS, Brazil}

\begin{abstract}
In this work we estimate the differential and total cross sections for the high energy deeply virtual Compton scattering in the weak sector. In the weak neutral sector one considers neutrino scattering off an unpolarized proton target through the exchange of $Z^0$. We numerically compute the process $Z^* p \rightarrow \gamma p$ within the QCD color dipole formalism, which successfully describes the current high energy electromagnetic DVCS experimental data. We also discuss possible applications for the weak charged sector and perform predictions for scattering on nuclear targets.

\end{abstract}

\maketitle

\section{Introduction}

The cross section of hard scattering processes can be
written as the convolution of parton distributions (PDFs) and the cross
sections of hard subprocesses computed at parton level using perturbative
QCD. The usual PDFs are the
diagonal element of an operator in the Wilson operator product expansion, which can be extracted from data for inclusive processes. On the other hand, there are a set of exclusive reactions which are
described by the off-diagonal elements of the density matrix, where the
momentum, helicity or charge of the outgoing target are not the same as
those of the correspondent ingoing particle. Examples of such reactions
are the virtual photon Compton scattering and the vector meson electroproduction. In these cases, the difference with the inclusive case is the
longitudinal components of the incoming and outgoing proton momentum.  Precision data are becoming available for hard scattering processes whose
description requires knowledge of these generalized parton
distributions (GPDs). 

In recent years, significant effort was made to access GPDs through the measurement of hard exclusive electroproduction processes. The simplest process in this respect is the electromagnetic deeply virtual Compton scattering (DVCS), that is an electron scatters off a nucleon producing a real photon in the final state $\gamma^*p\rightarrow \gamma p$. On the other hand, studies on exclusive processes mediated by weak interactions have increased motivated by experimental proposals for neutrino factories and high intensity neutrino beams facilities. Using a neutrino beam instead of an electromagnetic one the DVCS process into the weak sector can be investigated. For the neutral weak current one has the process $Z^0 p\rightarrow \gamma p$, whereas for the charged current case occurs the process $W^{\pm}p\rightarrow \gamma N$. In these cases, one expects sensitivity to a different flavor decomposition of GPDs and also due to the presence of the axial part of the V-A interaction they are sensitive to a different set of GPDs. Therefore, neutrino scattering on nucleons in the deeply virtual kinematics can shed light on the hadronic interactions mediated by electroweak currents.

 Recently, some sound works have addressed the DVCS process in case of weak interactions. In Ref. \cite{Amore:2004ng} the  neutrino scattering off protons in the deeply virtual kinematics, which describes under a unified formalism elastic and deep inelastic neutrino scattering,  was first studied  via an extension of the notion of DVCS to the case of a neutral current exchange. Afterwards, in Ref. \cite{Coriano:2004bk} the investigation was extended for both charged and the neutral electroweak sectors, where the structure of the leading twist amplitudes of the process has been discussed. In a very updated investigation \cite{PMR}, the DVCS in weak sector is considered in the generalized Bjorken limit and calculated at the leading twist within the framework of the nonlocal light-cone expansion via coordinate space QCD string operators. The cross sections for neutrino scattering off the nucleon, relevant for future high-intensity neutrino beam facilities are estimated. 

Besides the GPDs formalism for exclusive process, at high energies a simple and intuitive approach has attracted a wide interest. It is the color dipole formalism \cite{dipole}, where the interaction is described in the target rest frame and relies on the Fock states expansion of the projectile and their further scattering with the target. Specifically, the exclusive process $(\gamma, \,Z^0, \,W^{\pm})\,p\rightarrow E\,p$ is described  by the interaction of $q\bar{q}$ pairs (color dipoles), in which the virtual boson fluctuate into, with the nucleon. The scattering amplitude is given by the convolution of the dipole-nucleon cross section with the overlap of the virtual incoming boson and outgoing exclusive ($E$) final state wave functions. This approach provides a very good description of the data on $\gamma p$ inclusive production, $\gamma \gamma$ processes, diffractive deep inelastic and vector meson production (For a recent review, see \cite{fss}). In particular, the electromagnetic DVCS cross section is nicely reproduced in several implementations of the dipole cross section at low $x$ \cite{FM,FMS}. In particular, the ability to predict cross sections at very  high energies depends critically on the reliability of extrapolations from current measurements into regions of lower virtualities of the elementary scattering process, and much smaller values of the fractional momentum $x$ carried by the parton constituents of the hadrons. The color dipole approach is quite robust for doing this job.

In this letter we study the application of color dipole formalism to the DVCS process in electroweak sector. For real photon in the final state, the QCD calculation becomes not trivial and saturation models can provide a suitable scheme for dealing with low momenta scales appearing in the problem. The present calculation is strongly motivated by recent studies in Refs. \cite{Amore:2004ng,Coriano:2004bk,PMR}, where the leading twist amplitudes for deeply virtual neutrino scattering (DVNS) have been discussed. Those investigations have considered the generalized Bjorken limit within the framework of the nonlocal light-cone expansion, which involves GPDs only for quarks. On the other hand, at higher energies the processes induced by gluons are increasingly important. Therefore, the studies presented here are complementary to those in Refs. \cite{Amore:2004ng,Coriano:2004bk,PMR}. In next section, we make a short summary on the color dipole approach applied to electromagnetic DVCS and show its success in describing the high energy accelerator data.  In Sec.  3, we extend the formalism for the electroweak sector, considering for simplicity the neutral current case. Prediction are done for the differential cross section $d\sigma/dt$ and total cross sections considering a proton target. In addition, phenomenology considering nuclear targets is discussed.  Finally, in last section we summarize the results.

\section{High energy electromagnetic DVCS and the color dipole approach}

Let us introduce the main formulas concerning the color dipole picture applied to DVCS processes. In the dipole model \cite{dipole}, deep inelastic scattering is viewed as the interaction of a colour dipole, i.e. mostly a quark--antiquark pair, with the proton.  The transverse size of the pair is denoted by $\vec{r}$ and a quark carries a fraction $z$ of the photon's light-cone momentum.  In the proton rest frame, the dipole lifetime is much longer than the lifetime of its interaction with the target proton.  Therefore, the elastic $\gamma^* p$ scattering is assumed to proceed in three stages: first the incoming virtual photon fluctuates into a quark--antiquark pair, then the $q\bar{q}$ pair scatters elastically on the proton, and finally the $q\bar{q}$ pair recombines to form a virtual photon. The amplitude for the elastic process $\gamma^* p\rightarrow \gamma^* p$, $\mathcal{A}^{\gamma^*p}(x,Q,\Delta)$, is simply the product of amplitudes of these three subprocesses integrated over the dipole variables $\vec{r}$ and $z$:
\begin{eqnarray}
  \mathcal{A}^{\gamma^*p}(x,Q,\Delta) & = & \sum_f \sum_{h,\bar h} \int\!\dif^2\vec{r}\,\int_0^1\!\dif{z}\,\Psi^*_{h\bar h}(r,z,Q)\nonumber \\
& \times & \mathcal{A}_{q\bar q}(x,r,\Delta)\,\Psi_{h\bar h}(r,z,Q),
  \label{eq:elamp}
\end{eqnarray}
where $\Psi_{h\bar h}(r,z,Q)$ denotes the amplitude for the incoming virtual photon to fluctuate into a quark--antiquark dipole with helicities $h$ and $\bar h$ and flavour $f$. The quantity $\mathcal{A}_{q\bar q}(x,r,\Delta)$ is the elementary amplitude for the scattering of a dipole of size $\vec{r}$ on the proton, $\vec{\Delta}$ denotes the transverse momentum lost by the outgoing proton (with $t=-\Delta^2$), $x$ is the Bjorken variable and $Q^2$ is the photon virtuality. The elementary elastic amplitude $\mathcal{A}_{q\bar q}$ can be related to the $S$-matrix element $S(x,r,b)$ for the scattering of a dipole of size $\vec{r}$ at impact parameter $\vec{b}$ \cite{kowtea}:
\begin{eqnarray}
  \mathcal{A}_{q\bar q}(x,r,\Delta) & = & \int\!\dif^2\vec{b}\;\mathrm{e}^{-\mathrm{i}\vec{b}\cdot\vec{\Delta}}\,\mathcal{A}_{q\bar q}(x,r,b)\nonumber \\
  & = & \mathrm{i}\,\int \dif^2\vec{b}\;\mathrm{e}^{-\mathrm{i}\vec{b}\cdot\vec{\Delta}}\,2\left[1-S(x,r,b)\right]. 
  \label{eq:smatrix}
\end{eqnarray}
This corresponds to the notion of impact parameter when the dipole size is small compared to the size of the proton.  The integration over $\vec{b}$ of the S-matrix element motivates the definition of the $q\bar{q}$--$p$ differential cross section as
\begin{eqnarray}
  \frac{\dif\sigma_{q\bar q}}{\dif^2 \vec{b}}= 2[1-\mathrm{Re}\,S(x,r,b)].
  \label{eq:difsiggp}
\end{eqnarray}

The amplitude for production of the exclusive final state such as  a real photon in DVCS, is given by
\begin{eqnarray}
 & &  \mathcal{A}^{\gamma^* p\rightarrow \gamma p}(x,Q,\Delta) = \int\!\dif^2\vec{r}\int_0^1\!\dif{z}\;(\Psi_{\gamma}^{*}\Psi_{\gamma^*})_{T}\;\mathcal{A}_{q\bar q}(x,r,\Delta)\nonumber \\
 & & = \mathrm{i}\,\int\!\dif^2\vec{r}\int_0^1\!\dif{z}\int\!\dif^2\vec{b}\;(\Psi_{\gamma}^{*}\Psi_{\gamma^*})_{T}\;\mathrm{e}^{-\mathrm{i}\vec{b}\cdot\vec{\Delta}}\;\frac{\dif\sigma_{q\bar q}}{\dif^2 \vec{b}},
  \label{eq:ampvecm}
\end{eqnarray}
where $(\Psi^*\Psi)_{T}$ denotes the overlap of the vitual incoming photon and outgoing real photon wave functions.  For DVCS, the amplitude involves a sum over quark flavours. Here, one assumes that the size of the quark--antiquark pair is much smaller than the size of the proton. After considerations about the  non-forward wave functions we finally obtain \cite{kowtea},
\begin{eqnarray}
  \mathcal{A}^{\gamma^* p\rightarrow \gamma p}(x,Q,\Delta) & = & \mathrm{i}\,\int\!\dif^2\vec{r}\int_0^1\!\dif{z}\int\!\dif^2\vec{b}\;(\Psi_{\gamma}^{*}\Psi_{\gamma^*})_{T}\nonumber \\ 
 &\times & \mathrm{e}^{-\mathrm{i}[\vec{b}-(1-z)\vec{r}]\cdot\vec{\Delta}}
  \;\frac {\dif\sigma_{q\bar q}}{\dif^2\vec{b}}.
  \label{eq:newampvecm}
\end{eqnarray}
The elastic diffractive cross section is then given by
\begin{eqnarray}
  \frac{\dif\sigma^{\gamma^* p\rightarrow \gamma p}}{\dif t}
  & = & \frac{1}{16\pi}\left\lvert\mathcal{A}^{\gamma^* p\rightarrow \gamma p}(x,Q,\Delta)\right\rvert^2 
  \label{eq:xvecm1}
\end{eqnarray}

 For one has real photons in final state, only the transversely polarized overlap function contributes to the cross section.  Summed over the quark helicities, for a given quark flavour $f$ it is given by

\begin{eqnarray}
  (\Psi_{\gamma}^*\Psi_{\gamma^*})_{T}^f & = & \frac{N_c\,\alpha_{\mathrm{em}}e_f^2}{2\pi^2}\left\{\left[z^2+\bar{z}^2\right]\epsilon K_1(\epsilon r) m_f K_1(m_f r) \right.\nonumber \\
& + &  \left. m_f^2 K_0(\epsilon r) K_0(m_f r)\right\},
  \label{eq:overlap_dvcs}
\end{eqnarray}
where we have defined the quantities $\varepsilon^2 = z\bar{z}\,Q^2+m_f^2$ and $\bar{z}=(1-z)$.

\begin{figure}[t]
\includegraphics[scale=0.47]{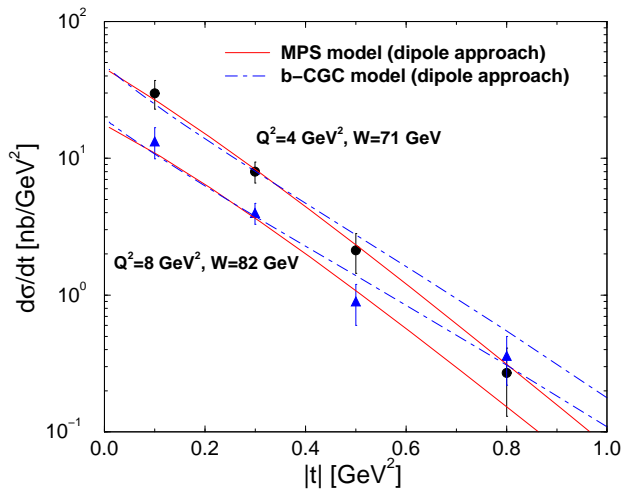}
\caption{(Color online) Differential cross section $d\sigma/dt$ for DVCS on a proton as a function $|t|$ at fixed energies and $Q^2$. Experimental data from DESY-HERA \cite{Aktas:2005ty}  (see text).}
\label{fig:1}
\end{figure}

Before discussing the specific model for the dipole-cross section some corrections to the exclusive processes should be shortly addressed. For DVCS one  should use the off-diagonal (or generalized) gluon distribution, since the exchanged gluons carry different fractions $x$ and $x^\prime$ of the proton's (light-cone) momentum. The skewed effect can be accounted for, in the limit that $x^\prime \ll x \ll 1$, by multiplying the elastic differential cross section by a factor $R_g$, given by \cite{Shuvaev:1999ce}
\begin{eqnarray} 
\label{eq:Rg}
  R_g(\lambda) = \frac{2^{2\lambda+3}}{\sqrt{\pi}}\frac{\Gamma(\lambda+5/2)}{\Gamma(\lambda+4)},\! \quad\text{with}\!\quad \lambda \equiv \frac{\partial\ln\left[\mathcal{A}(x,Q^2,|t|)\right]}{\partial\ln(1/x)}.\nonumber
\end{eqnarray}

For phenomenology on dipole cross section, we take two saturation models which successfully describe the high energy DVCS data. The first one is the impact parameter saturation model \cite{KMW} (hereafter b-CGC). The idea behind it is to introduce the impact parameter dependence into the CGC model \cite{Iancu:2003ge}, where now the $q\bar{q}-p$ differential cross section is given by $\frac{\dif\sigma_{q\bar{q}}}{\dif^2\vec{b}}= 2\,N(x,\,r;\,b)$, where 
\begin{equation} 
\label{eq:bcgc}
 N(x,\,r\,;\,b) =\begin{cases}
  \mathcal{N}_0\left(\frac{rQ_{\mathrm{sat}}}{2}\right)^{2\left(\gamma_s+\frac{1}{\kappa\lambda Y}\ln\frac{2}{rQ_{\mathrm{sat}}}\right)} & :\quad rQ_{\mathrm{sat}}\le 2\\
  1-\mathrm{e}^{-A\ln^2(BrQ_{\mathrm{sat}})} & :\quad rQ_{\mathrm{sat}}>2
  \end{cases},
\end{equation}
where $Y=\ln(1/x)$ and $Q_{\mathrm{sat}}(x,b)$ is the impact parameter saturation scale defined as \cite{KMW}:
\begin{equation} 
\label{eq:bcgc1}
Q_{\mathrm{sat}}(x,b)=\left(\frac{x_0}{x}\right)^{\frac{\lambda}{2}}\;\left[\exp\left(-\frac{b^2}{2B_{\rm CGC}}\right)\right]^{\frac{1}{2\gamma_s}}.
\end{equation}

 In the b-CGC model the evolution effects are included via an approximate solution to the Balitsky--Kovchegov equation \cite{BK}. The parameters of the model can be obtained in Table 4 of Ref. \cite{KMW}. In particular, one has the slope $B_{\mathrm{CGC}}=5.5$ GeV$^{-2}$. The constants $A$ and $B$ are obtained from continuity conditions at  $rQ_{\mathrm{sat}}=2$. 

\begin{figure}[t]
\includegraphics[scale=0.47]{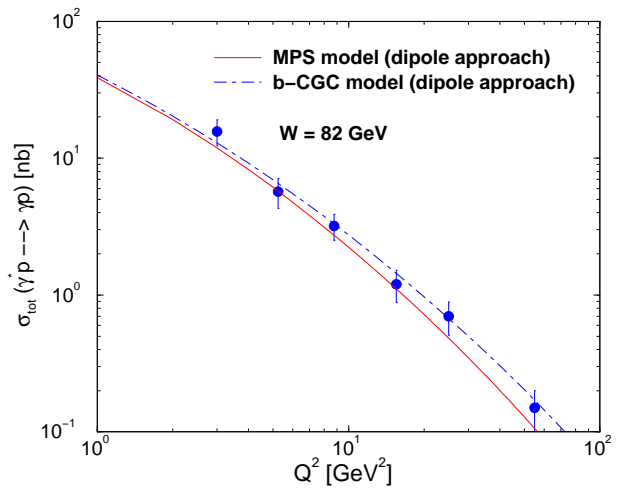}
\caption{(Color online) The total cross section for DVCS on a proton as a function $Q^2$ at fixed energy $W=82$ GeV. Experimental data from DESY-HERA \cite{Aktas:2005ty}  (see text).}
\label{fig:1b}
\end{figure}
  
The second saturation model that we will consider is that one presented in Ref. \cite{MPS} (hereafter MPS model). It has the great advantage of giving directly the $t$ dependence of elastic differential cross section without the necessity of considerations about the impact parameter details of the process. The authors expressed the exclusive production cross-sections in the high-energy limit in terms of the Fourier-transformed dipole scattering amplitude off the proton $\mathcal{A}_{q\bar q}(x,r,\Delta)$.  An important result about the growth of the dipole amplitude towards the saturation regime is the geometric scaling regime \cite{gsincl,travwaves}. It first appeared in the context of the proton structure function, which involves the
dipole scattering amplitude at zero momentum transfer. At small values of $x,$
instead of being a function of both the variables $r$
and $x$, the dipole scattering amplitude is actually a function of the single
variable $r^2Q^2_{\mathrm{sat}}(x)$ up to inverse dipole sizes significantly larger than the
saturation scale $Q_{\mathrm{sat}}(x).$ More precisely, one can write
\begin{eqnarray}
\label{eq:geomsc0}
\mathcal{A}_{q\bar q}(x,r,\Delta=0)=2\pi R_p^2\: N\left(r^2Q_{\mathrm{sat}}^2(x)\right)\ ,
\end{eqnarray}
implying the geometric scaling of the total cross-section
at small $x$, i.e. $\sigma^{\gamma^*p\rightarrow X}_{\text{tot}}(x,Q^2)=
\sigma^{\gamma^*p\rightarrow X}_{\text{tot}}(\tau_p\!=\!Q^2/Q_{\mathrm{sat}}^2).$ This geometric scaling property can be extended to the case
of non zero momentum transfer \cite{MAPESO}, provided $r\Delta\ll 1$. Therefore, it has been obtained that
equation \eqref{eq:geomsc0} can be generalized to
\begin{eqnarray}
\mathcal{A}_{q\bar q}(x,r,\Delta)= 2\pi R_p^2 \:F(\Delta)\:
N(r^2Q_{\mathrm{sat}}^2(x,\Delta))\ ,
\end{eqnarray}
with the asymptotic behaviors $Q_{\mathrm{sat}}^2(x,\Delta)\sim
\max(Q_0^2,\Delta^2)\,\exp[-\lambda \ln(x)]$ and an unknown form factor
$F(\Delta)$ of non-perturbative origin. Specifically, the $t$ dependence of the saturation scale is parametrised as
\begin{eqnarray}
\label{qsatt}
Q_{\mathrm{sat}}^2\,(x,|t|)=Q_0^2(1+c|t|)\:\left(\frac{1}{x}\right)^{\lambda}\,, \end{eqnarray}
in order to interpolate smoothly between the small and intermediate transfer
regions. The form factor $F(\Delta)=\exp(-B|t|)$ catches the transfer dependence of the proton vertex, which is factorised from the
projectile vertices and  does not spoil the geometric scaling properties. Finally, the scaling function $N$ is obtained from the forward saturation model
\cite{Iancu:2003ge}, whose functional form is given by Eq. (\ref{eq:bcgc}). The parameters for $N$ are taken from the original CGC model \cite{Iancu:2003ge}. The final expression of MPS saturation model is given by
\begin{eqnarray}
\label{sigdipt}
\mathcal{A}_{q\bar q}(x,r,\Delta)= 2\pi R_p^2\,e^{-B|t|}N \left(rQ_{\mathrm{sat}}(x,|t|),x\right),
\end{eqnarray}
which is an extension of the forward model \cite{Iancu:2003ge} including the QCD
predictions for non zero momentum transfer, which reproduces the
initial model for $|t|=0$ and ensures that the saturation scale has the
correct asymptotic behaviors. The remaining parameters are $c=3.776$ GeV$^{-2}$ and $B=3.740$ GeV$^{-2}$ (obtained from a fit with boosted-Gaussian meson wavefunction to describe exclusive vector meson production). In order to compute the DVCS cross section, the amplitude above can be directly replaced in first line of Eq (\ref{eq:ampvecm}).

In order to show the reliability of the models mentioned above, we compare them to the recent high energy DVCS measurements from DESY-HERA. In Fig. \ref{fig:1} the differential cross section $d\sigma/dt$ is shown for distinct energies and virtualities. The solid curves represent the results for MPS model whereas the dot-dashed ones represent the b-CGC model. Both models are in agreement with data \cite{Aktas:2005ty} (only systematic errors are presented), presenting different trends towards large $t$. This is due to the different implementation for the $t$ dependence of the elementary elastic amplitude $\mathcal{A}_{q\bar q}(x,r,\Delta)$. A detailed analysis about the analytical/numerical results of the distinct models is beyond the scope of the present study. Finally, in Fig. \ref{fig:1b} the total cross section is presented as a function of $Q^2$ at fixed energy $W=82$ GeV, including the experimental results \cite{Aktas:2005ty}. The agreement is once again is very good. In this case, the numerical results for both models are obtained by integrating the differential cross section $d\sigma/dt$ for $|t|\leq 1$, which is the experimental cut.

\section{Phenomenological results for electroweak DVCS}

In this section we will extend the study for the case of electroweak DVCS, focusing first on the neutral weak current. This can occur for instance in the neutrino scattering off an unpolarized proton target through the exchange of $Z^0$ instead a photon as in the electromagnetic case. In addition, the virtual exchange can be a $W^{\pm}$ with the production of an energetic photon, a $\mu^{\pm}$, with either a recoiling nucleon or nucleon resonance. Using the color dipole picture, the elastic cross section for the high energy neutral weak DVCS scattering on a proton target can be written as
\begin{eqnarray}
 & &  \mathcal{A}^{Z^* p\rightarrow \gamma p}(x,Q,\Delta) = \int\!\dif^2\vec{r}\int_0^1\!\dif{z}\;(\Psi_{\gamma}^{*}\Psi_{Z^0})_{T}\;\mathcal{A}_{q\bar q}(x,r,\Delta)\nonumber \\
 & & = \mathrm{i}\,\int\!\dif^2\vec{r}\int_0^1\!\dif{z}\int\!\dif^2\vec{b}\;(\Psi_{\gamma}^{*}\Psi_{Z^0})_{T}\;\mathrm{e}^{-\mathrm{i}\vec{b}\cdot\vec{\Delta}}\;\frac{\dif\sigma_{q\bar q}}{\dif^2 \vec{b}},
  \label{eq:ampweak}
\end{eqnarray}
where now $(\Psi_{\gamma}^*\Psi_{Z^0})_{T}$ denotes the overlap of the incoming $Z^0$ and outgoing real photon final state wave functions. Explicit expressions for the boson wavefunctions can be found in Ref. \cite{ZOLLER1,ZOLLER2}.  Corrections for skewness is implicitly understood, as done in previous section. The present formalism has been successfully used in calculations of neutral/charged current neutrino cross sections \cite{KUTAK,MPRD1,MPRD2} and neutrino-nucleon structure functions \cite{MPLB,ZOLLER1}. For the charged current case, the overlap of the wave functions $(\Psi_{\gamma}^{*}\Psi_{Z^0})$ is replaced by $(\Psi_{\gamma}^{*}\Psi_{W^{\pm}})$, with a possible correction for the corresponding production of nucleon resonance.

\begin{figure}[t]
\includegraphics[scale=0.47]{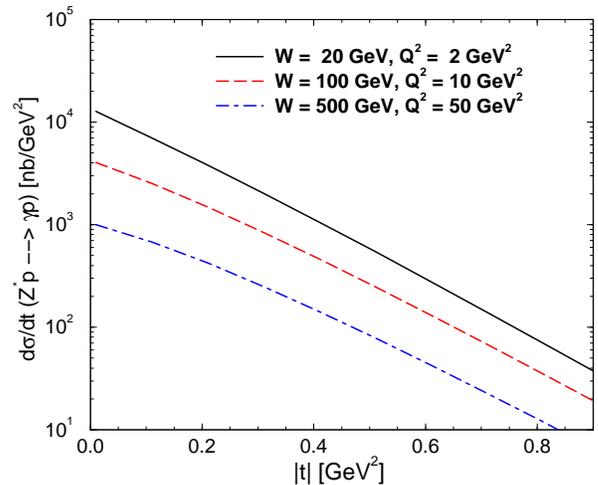}
\caption{(Color online) The differential cross section $d\sigma /dt$ for the weak neutral DVCS process (see text).}
\label{fig:2}
\end{figure}

We are now ready to compute the numerical results. For our purpose here, the real part of amplitude will be disregarded. For not so large $Q^2$, the effective power $\lambda (Q^2) \approx 1.26$ and the real part correction $\eta \simeq \tan(\pi \lambda/2)$ contributes just a few percents. In Fig. \ref{fig:2} the differential cross section $d\sigma /dt$ is shown as a function of $|t|$ using expression on Eq. (\ref{eq:ampweak}). In order to investigate the sensitivity on energy and virtuality, tree representative cases are considered. We start with a low energy, $W=20$ GeV and small $Q^2=2$ GeV$^2$, which is suitable for the neutrino beam facilities. The intermediate case is compatible with the DESY HERA kinematical range for DVCS. Finally, we extrapolate the results for very higher energy $W=500$ GeV and large $Q^2=50$ GeV$^2$. At small $|t|\leq 0.4$ the differential cross section is consistent with an exponential behavior and a $Q^2$ dependent slope. However, the general $|t|$ dependence is determined by the dipole cross section, $d\sigma_{q\bar{q}}/dt$. We checked that the cross section can be parameterized as $d\sigma/dt=a\,(1+c\,|t|)\exp(-B|t|)$, with the constants $a,\,c$ and $B$ depending on $Q^2$.

\begin{figure}[t]
\includegraphics[scale=0.47]{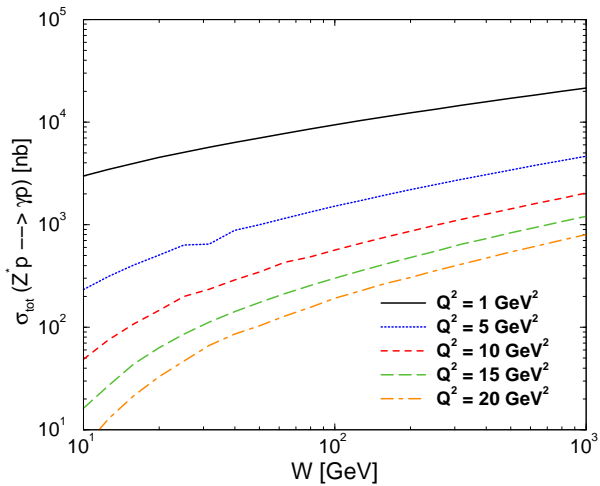}
\caption{(Color online) The total cross section for weak neutral DVCS as a function of energy at fixed virtualities (see text).}
\label{fig:3}
\end{figure}

In the following we compute the total cross section, performing the integration over $|t|\leq 1$ GeV$^2$. In Fig. \ref{fig:3} the energy dependence is presented for $W_{Zp} \geq 10$ GeV at fixed virtualities $Q^2=1\!-\!20$ GeV$^2$. Roughly speaking, this interval probes values of $x$ going from $x\approx 10^{-1}$ to $x\approx 10^{-6}$.  The energy power of the cross section, $\sigma_{tot}\propto W^{\alpha}$, ranges on $\alpha = 0.38-0.65$, which increases with $Q^2$.  For instance, in high energies one has $\sigma_{tot}(Z^* p \rightarrow \gamma p)= 1.55\,\mathrm{\mu b}\,(W_{Zp}/\mathrm{GeV})^{0.38}$ at $Q^2=1$ GeV$^2$ and $\sigma_{tot}(Z^*p \rightarrow \gamma p)= 9\,\mathrm{nb}\,(W_{Zp}/\mathrm{GeV})^{0.65}$ at $Q^2=20$ GeV$^2$. Therefore, the cross sections are large and a factor $10^2$ higher compared to the DVCS results measured at HERA. As a short comment, the effective power $\alpha$ decreases towards low virtualities as a consequence of the non-linear effects in the dipole cross section since the saturation scale, $Q_{\mathrm{sat}}(x)=Q_0\exp[-0.5\lambda\ln(x)]$,  is of order 1 GeV or even larger in the kinematical range considered here.

\begin{figure}[t]
\includegraphics[scale=0.47]{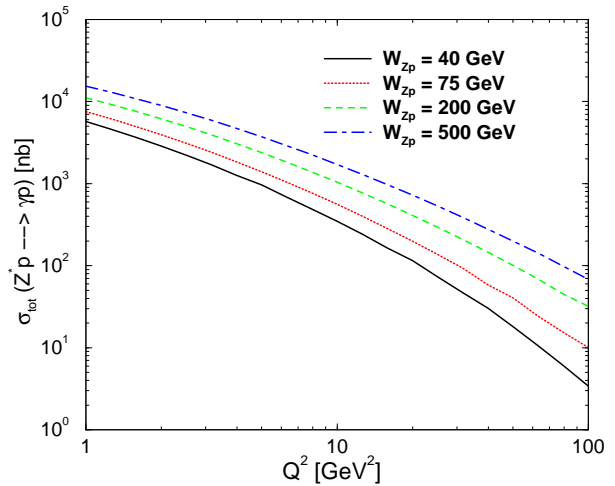}
\caption{(Color online)  The total cross section for weak neutral DVCS as a function of $Q^2$ at fixed energy (see text).}
\label{fig:4}
\end{figure}

In Fig. \ref{fig:4} the total cross section is shown as a function of $Q^2$ at fixed energies, 40 GeV $\leq W_{Zp} \leq$ 500 GeV. The virtuality dependence is similar to that appearing on DVCS results at HERA. Once again, the absolute values are large at high energies reaching to $\sigma_{tot}\simeq 10$ $\mu$b for $Q^2=1$ GeV$^2$. In the interval $1\leq Q^2 \leq 10$ GeV$^2$ the cross section can be parameterized as $\sigma_{tot}=a/(Q^2)^{\delta}$, with the constants $a$ and $\delta \approx 1$ depending on $W$. When going out of this interval, the $Q^2$ dependence is steeper and more involved. For example, one obtains $\sigma_{tot}(Z^* p \rightarrow \gamma p)= 5.9\,\mathrm{\mu b}/(Q_0^2/Q^2)^{1.1}$ for $W_{Zp}= 40$ GeV and $\sigma_{tot}(Z^*p \rightarrow \gamma p)= 11\,\mathrm{\mu b}\,(Q_0^2/Q^2)^{0.97}$ at $W_{Zp}=200$ GeV, where $Q_0^2 =1$ GeV$^2$.

The present calculation can be serve as an input for electroweak DVCS in neutrino interactions. In general, a neutrino induced DVCS process in a nucleon is given by the reaction $l_1(k)+N_1(p_1)\rightarrow l_2(k^{\prime})+N_2(p_2)+\gamma(q_2)$, where a neutrino $l_1$ scatters from a nucleon $N_1$ to a final state $l_2$, nucleon $N_2$ and real photon $\gamma$. The differential cross section in the target rest frame, in which the weak virtual boson four-momentum $q$ has no transverse components, assumes the form \cite{Amore:2004ng,Coriano:2004bk,PMR},
\begin{eqnarray}
\frac{d^4\sigma}{dxdQ^2d|t|d\phi}= \frac{1}{64\,s}\,\frac{1}{(2\pi)^4}\,\frac{1+x(M/E_{\nu})}{ME_{\nu}x\,\left[y+2x(M/E_{\nu})\right]^2}\,\left|T\right|^2,\nonumber
\end{eqnarray}
where $T\propto \mathcal{A}(x,r,\Delta)$ represents the invariant matrix element. The invariants in equation above are given by $s\equiv (k+p_1)^2$,  $Q^2\equiv -q^2$ and $y\equiv (p_1\cdot q)/(p_1\cdot k)$. The quantity $E_{\nu}$ denotes the energy of the incoming neutrino beam and $\phi$ the angle between the lepton and nucleon scattering planes. In electromagnetic DVCS, $T$ receives a large contribution from the Bethe-Heitler (BH) process. Considering a neutrino scattering off an unpolarized proton target through the exchange of $Z^0$, one only measures the Compton contribution since the photon can not be emmited by the neutrino. In the case of weak charged current case, where neutrino-nucleon scattering occurs through the exchange of $W^{+}$ with a proton in the final state, one has both DVCS and BH contributions since the outgoing muon can emit the real photon.

Finally, we will present a phenomenological investigation on the weak DVCS process for  neutrino scattering off nucleus. In order to do so, we will rely on the geometric scaling property \cite{gsincl,travwaves} of the saturation models within the color dipole approach. The geometric scaling  means that  the total $\gamma^* p$
cross section at large energies is not a function of the two
independent variables $x$ and $Q$, but is rather a function of the
single variable $\tau_p = Q^2/Q_{\mathrm{sat}}^2(x)$ as shown
in Ref. \cite{gsincl}. That is, $\sigma_{\gamma^*p}(x,Q^2)=\sigma_{\gamma^*p}(\tau_p)$. In Refs. \cite{travwaves} it was shown that the geometric scaling observed in experimental data can be understood theoretically in the context of non-linear QCD evolution with fixed and running coupling. Recently, the high energy $l^{\pm}p$,  $pA$ and $AA$ collisions have been related through geometric scaling \cite{Armesto_scal}. Within the color dipole picture and making use of a rescaling of the impact parameter of the $\gamma^*h$ cross section in terms of hadronic target radius $R_h$, the nuclear dependence of the $\gamma^*A$ cross section is absorbed in the $A$-dependence of the saturation scale via geometric scaling. The relation reads as $\sigma^{\gamma^*A}_{tot}(x,Q^2)  =  \kappa_A\,\sigma^{\gamma^*p}_{tot}\,(Q_{\mathrm{sat},p} \rightarrow Q_{\mathrm{sat},A})$, where $\kappa_A = (R_A/R_p)^2$. The nuclear saturation scale was assumed to rise with the quotient of the transverse parton densities to the power $\Delta \approx 1$ and $R_A$ is the nuclear radius, $Q_{\mathrm{sat},A}^2=(A/\kappa_A)^{\Delta}\,Q_{\mathrm{sat},p}^2$. This assumption successfully describes small-$x$ data for $ep$ and $eA$ scattering using $\Delta =1.26$ and a same scaling curve for the proton and nucleus \cite{Armesto_scal}.

\begin{figure}[t]
\includegraphics[scale=0.47]{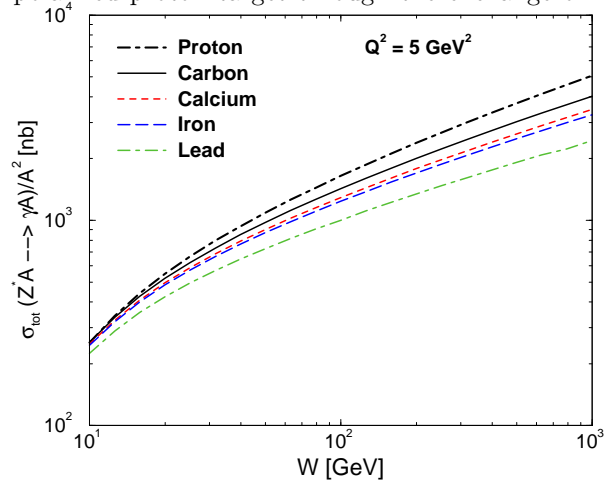}
\caption{(Color online)  The nuclear cross section for weak neutral DVCS as a function energy at fixed $Q^2=5$ GeV$^2$.  Results for representative nuclei are presented. (see text).}
\label{fig:5}
\end{figure}

Following the simple arguments of scaling, we replace $R_p \rightarrow R_A$ in Eq. (\ref{sigdipt}) and $Q_{\mathrm{sat},p}^2 \rightarrow (AR_p^2/R_A^2)^{\Delta}\,Q_{\mathrm{sat},p}^2$ in Eq. (\ref{qsatt}). In case of $\Delta=1$ the previous replacement becomes the usual assumption for the nuclear saturation scale, $Q_{\mathrm{sat},A}^2=A^{1/3}\,Q_{\mathrm{sat},p}^2$. The corresponding phenomenological results are presented in Fig. \ref{fig:5}. As a short comment, the present estimation corresponds to the coherent electroweak DVCS process, $Z^0A\rightarrow \gamma A\,X$, where the nucleus remains intact after interaction \cite{Pire}. Future studies should include the incoherent contribution, which is sizeable for high energies and heavy nuclei \cite{Pire}. The total cross section as a function of energy is shown for fixed virtuality $Q^2=5$ GeV$^2$ and representative nuclei. The effective power on energy is dependent on the mass number. For light nuclei like carbon the power follows similar trend as for the nucleon case but there is a clear suppression at high energies. For large nucleus, i.e. Ca, Fe or Pb, the effective power systematically diminishes due to nuclear shadowing. In the saturation model, the enhancement of the nuclear saturation scale due to the presence of nuclear targets leads to an additional flattening of the energy growth. The nuclear suppression as a function of energy is verified by contrasting the nuclear results to the proton case (upper dot-dashed curve). In order to quantify the size of nuclear suppression, let us give some examples. For carbon it is of order 15\% at high energies, whereas for lead it reaches 50 \%. The present study for nuclear targets should be timely for future neutrino-nucleus experiments as Minerva \cite{minerva}.

\section{Summary}

 We estimate the differential $d\sigma/dt$ and total cross section for the high energy electroweak deeply virtual Compton scattering on a proton target and extended for nuclear targets. We take into account the QCD color dipole formalism, which is quite successful in describing small-$x$ data.  The absolute values for  the cross sections are large, higher than electromagnetic DVCS measured at DESY-HERA, which makes its measurement feasible. The main theoretical  uncertainty is the specific model for the dipole cross section. We have considered two saturation model, which are able to correctly describe electroweak DVCS data. Skewedness correction have been properly included in calculations, giving a sizable enhancement.  In the present study, one considers neutrino scattering off an unpolarized proton target through the exchange of $Z^0$, that is neutral current case. The extension for charged weak current is possible within the dipole approach.

\begin{acknowledgments}
This work was financed by the Brazilian funding
agency CNPq. The author is grateful to L. Motyka, G. Watt and C. Marquet for useful informations. The clarifying comments of V. Zoller and B. Pire are acknowledged.
\end{acknowledgments}


\begin{thebibliography}{99}

\bibitem{Amore:2004ng}
  P.~Amore, C.~Coriano and M.~Guzzi, JHEP {\bf 0502}, 038 (2005).

\bibitem{Coriano:2004bk}
  C.~Coriano and M.~Guzzi,  Phys.\ Rev.\  D {\bf 71}, 053002 (2005).

\bibitem{PMR}
A. Psaker, W. Melnitchouk and A. V. Radyushkin, Phys. Rev. D 75, 054001 (2007).

\bibitem{dipole}  N.~N.~Nikolaev and B.~G.~Zakharov, Z. Phys. {\bf  C49}, 607
(1991); Z. Phys. {\bf C53}, 331 (1992); A.~H.~Mueller, Nucl. Phys.
{\bf B415}, 373 (1994); A.~H.~Mueller and B.~Patel, Nucl. Phys. {\bf B425}, 471
(1994).

\bibitem{fss}
J.~R.~Forshaw, R.~Sandapen and G.~Shaw,
JHEP {\bf 0611}, 025 (2006).

\bibitem{FM}
L.~Favart and M.~V.~T.~Machado,
Eur.\ Phys.\ J.\ C {\bf 34}, 429 (2004).

\bibitem{FMS}
L.~Favart, M.~V.~T.~Machado and L. Schoeffel, arXiv:hep-ph/0511069.

\bibitem{kowtea}
H. Kowalski and D. Teaney,
{\it Phys. Rev.} {\bf D68} (2003) 114005.

\bibitem{Shuvaev:1999ce}
A.~G.~Shuvaev {\it et al.},
Phys.\ Rev.\ D {\bf 60}, 014015 (1999).

\bibitem{KMW}
H. Kowalski, L. Motyka and G. Watt,
{\it Phys. Rev.} {\bf D74} (2006) 074016.

\bibitem{Iancu:2003ge}
  E.~Iancu, K.~Itakura and S.~Munier,
  Phys.\ Lett.\ B {\bf 590} (2004) 199.

\bibitem{BK}
I.~Balitsky,
Nucl.\ Phys.\ B {\bf 463}, 99 (1996);
Y.~V.~Kovchegov,
Phys.\ Rev.\ D {\bf 60}, 034008 (1999);
Phys.\ Rev.\ D {\bf 61}, 074018 (2000).

\bibitem{MPS}
C. Marquet, R. Peschanski and G. Soyez, arXiv:hep-ph/0702171.

\bibitem{gsincl}
A.M. Stasto, K. Golec-Biernat and J. Kwiecinski,
{\it Phys. Rev. Lett.} {\bf 86} (2001) 596.

\bibitem{travwaves}
S. Munier and R. Peschanski,
{\it Phys. Rev. Lett.} {\bf 91} (2003) 232001;
{\it Phys. Rev.} {\bf D69} (2004) 034008;
{\bf D70} (2004) 077503.

\bibitem{MAPESO}
C. Marquet, R. Peschanski and G. Soyez,
{\it Nucl. Phys.} {\bf A756} (2005) 399.

\bibitem{Aktas:2005ty}
  A.~Aktas {\it et al.}  [H1 Collaboration],
  Eur.\ Phys.\ J.\ C {\bf 44} (2005) 1.
\bibitem{ZOLLER1} R. Fiore and V.R. Zoller, JETP Lett. {\bf 82} (2005) 385.

\bibitem{ZOLLER2} R. Fiore and V.R. Zoller, Phys. Lett. {\bf B632} (2006) 87.

\bibitem{KUTAK} K. Kutak and J. Kwieci\'nski,  Eur. Phys. J. C {\bf 29}, 521 (2003). 

\bibitem{MPRD1} M.V.T. Machado, Phys. Rev. {\bf D71} (2005) 114009.

\bibitem{MPRD2} M.V.T. Machado, Phys. Rev. {\bf D70} (2004) 053008.

\bibitem{MPLB}
M.B. Gay Ducati, M.M. Machado and M.V.T. Machado, Phys. Lett. B {\bf 644 }, 340 (2007). 

\bibitem{Armesto_scal}
  N.~Armesto, C.~A.~Salgado and U.~A.~Wiedemann,
  Phys.\ Rev.\ Lett.\  {\bf 94}, 022002 (2005).

\bibitem{minerva} D. Drakoulakos et al. [Minerva Collaboration], FERMILAB-PROPOSAL-0938; arXiv:hep-ex/0405002.

\bibitem{Pire} E.R. Berger, F. Cano, M. Diehl and B. Pire, Phys.\ Rev.\ Lett.\  {\bf 87}, 142302 (2001); F. Cano and B. Pire, Eur. Phys. J. A {\bf 19}, 423 (2004). 


\end{thebibliography}
\end{document}